\documentclass[letter]{aa}  

\usepackage{graphicx}
\usepackage{txfonts}
\usepackage{amsmath,amssymb}
\usepackage{longtable}
\usepackage{natbib}

\begin{document}

   \title{Orbital evolution of Saturn'{}s satellites due to the interaction between the moons and massive Saturn'{}s rings}
   \titlerunning{Interaction between the moons and Saturn's massive rings}

   %\subtitle{}

   \author{Ayano Nakajima
          \inst{1}
          \and
          Shigeru Ida\inst{1,2}
          \and
          Yota Ishigaki\inst{3,4}
          }

   \institute{Department of Earth and Planetary Sciences, Tokyo Institute of Technology, Tokyo 152-8551, Japan
              \\
              \email{nakajima.a.ah@m.titech.ac.jp}
         \and
             Earth-Life Science Institute, Tokyo Institute of Technology, Tokyo, Japan
             \\
             \email{ida@elsi.ac.jp}
             %\thanks{The university of heaven temporarily does not accept e-mails}
          \and
             Department of Earth and Planetary Sciences, University of Tokyo, Tokyo, Japan
           \and
             ISAS, JAXA, Kanagawa, Japan
             \\
             \email{y.ishigaki@stp.isas.jaxa.jp}
             }

  \abstract
   {Saturn'{}s mid-sized moons (satellites)
   have a puzzling orbital configuration with trapping in mean-motion resonances with every other pairs 
   (Mimas-Tethys 4:2 and Enceladus-Dione 2:1). 
   To reproduce their current orbital configuration
   on the basis of Crida \& Charnoz'{}s model of satellite formation from a hypothetical  ancient massive rings, adjacent pairs must pass 1st-order
   mean-motion resonances without being trapped. 
    }
   {
   The trapping could be avoided by fast orbital migration and/or excitation of the satellite'{}s eccentricity 
   caused by gravitational interactions between the satellites and the rings (the disk), which are still unknown. In our research, 
   we investigate the satellite orbital evolution due to interactions with the disk through full N-body simulations.
}
   {We performed global high-resolution N-body simulations of a self-gravitating particle disk interacting with a single satellite. 
   We used $N\sim10^5$ particles for the disk. 
   Gravitational forces of all the particles and their inelastic collisions are taken into account.     
   }  
   {Dense short-wavelength wake structure is created by the disk self-gravity 
   and global spiral arms with $m \sim$ a few is induced by the satellite.
   The self-gravity wakes regulate the orbital evolution of the satellite,
   which has been considered as a disk spreading mechanism but not as a driver for the orbital evolution.
   }
   {The self-gravity wake torque to the satellite is so effective 
   that the satellite migration is much faster than that was predicted with the spiral arms torque. 
   It provides a possible model to avoid the resonance capture of adjacent satellite pairs and 
   establish the current orbital configuration of Saturn'{}s mid-sized satellites. }


   \maketitle
%
%-------------------------------------------------------------------

\section{Introduction}
The orbital configuration of Saturn'{}s mid-sized moons, Mimas, Enceladus, Tethys, Dione and Rhea from inner to outer orbits, is puzzling:
they are trapped in mean-motion resonances 
for every other pairs (Mimas-Tethys 4:2 and Enceladus-Dione 2:1), but not for adjacent pairs.
The observed current fast tidal orbital expansion rate \citep{Lainey2012, Lainey2017} and  observations of the rings by Cassini suggest late formation of the satellites \citep[see e.g.][]{Ida2019} such as the formation model from 
a hypothetical ancient massive rings \citep{Charnoz2011,Crida2012}, which we refer to as the ``disk." 
Note, however, that a possible slower tidal orbital expansion rate in the past, before resonance locking between the orbital frequency and the planetary internal oscillation mode, could allow the satellite formation in the circumplanetary disk 4.5 G years ago \citep{Lainey2020}.

In the model of the formation from the disk, satellites were formed one after another
at the disk outer edge
and the outward orbital migrations of adjacent pairs of satellites 
due to planetary tide are generally convergent.
In that case, the satellite pairs are usually captured into a mutual 1st order mean-motion resonance, which is inconsistent with the current orbital configuration \citep[e.g.][]{Nakajima2019}.
The avoidance of such resonance capture requires moderate orbital eccentricity of the satellites
or fast orbital migration with the timescale smaller than the resonant libration period   \citep{Malhotra1996}.

Recent Cassini's observations determined 
the current rings mass as $M_{\rm {disk}} = (1.54 \pm 0.49) \times 10^{19} \simeq 0.4 \times$ Mimas mass \citep{Iess2019}. 
The rings still undergo viscous spreading and should have been much more massive in the past \citep{Salmon2010}.
\citet{Crida2012} suggested that the satellite-disk (rings) interaction is more effective 
for the orbital migration than the Saturn's tide until the satellite
reaches the 2:1 resonance with the disk outer edge, beyond which the disk torque would quickly decay.
They applied the theoretical model for a planet in a gap of 
a protoplanetary disk \citep[e.g.][]{Lin1986}
to estimate the migration rate of the satellite.
The satellite-disk interactions can also excite
the orbital eccentricity of the satellite \citep{Goldreich2003b, Duffell2015}.
The eccentricity excitation may be much faster than the eccentricity damping by the satellite's tide,
near the disk edge, as we will show in Section 2.
If the eccentricity is excited beyond a critical value,
the satellites can avoid the resonance capture to reach the current orbital configuration \citep{Nakajima2019}.
 
Previous studies on planet-disk interactions
usually assumed protoplanetary gas disks that are stable against self-gravitational instability, while
the rings are often in a marginally unstable state \citep[e.g.][]{Salo1995,Daisaka2001}.
Thus, it is important to investigate the interactions of a satellite and
a marginally unstable self-gravitating particle disk (rings) by high resolution N-body simulations.  

\citet{Hyodo2015} performed high resolution N-body simulations ($N=3\times10^4 - 5 \times 10^4$) 
of the formation of satellites from a disk with mass
$M_{\rm disk} \simeq (0.01$--$0.06) \, M_{\rm p}$ ($M_{\rm p}$ is the planet mass)
to find the dependence of the forming satellite mass on the initial disk mass.
On the other hand, they were not concerned with the orbital evolution of satellites and disk structures.

Here, we focus on the detailed evolution of the disk structures and satellite's orbit.
In our study, we perform high-resolution ($N\sim10^5$) N-body simulation of
particle disk evolution
due to the mutual gravitational interactions and inelastic collisions of the particles
and the disk's interactions with a satellite in an orbit exterior to the disk,
to investigate the orbital evolution of the satellite. 
   
\section{Methods}

We use a new N-body simulation code, ``GPLUM" (Ishigaki in prep),
that adopts the particle-particle particle-tree scheme \citep[P$^3$T,][]{Ohsino2011} for planetary formation. The P$^3$T scheme uses the fourth-order Hermite integrator to calculate gravitational interactions between particles within a cut-off radius
and the Barnes-Hut tree method for gravity from particles beyond the cut-off \citep{Iwasawa2016}, which guarantees higher-order integrations for close interactions and fast integrations for perturbations from a large number of distant particles simultaneously.
GPLUM adopts individual cut-off radius scheme for individual particles, depending on their mass and distance from the central star, resulting in a significant speedup of calculations, while keeping the accuracy.

We follow the orbital evolution of the satellite and the disk particles
by the gravitational interactions, inelastic collisions between the particles, 
and the accretion of the particles onto the satellite and the host planet.
When physical sizes overlap, we regard that a collision occurs.
For collisions between the disk particles, we apply inelastic collisions with the normal restitution coefficient $\epsilon_{\rm n} = 0.1$
and the tangential restitution coefficient $\epsilon_{\rm t} = 1$ (free-slip condition). 
When a particle collides with the satellite at the orbital radius larger than the Roche limit radius (Eq. (\ref{roche})), the collision results in gravitational binding of the particle and the satellite. Because in the results we show here, the satellite does not reenter the Roche limit,
we make a merged body from the satellite and the particle, 
keeping their total mass and momentum.

The particles are initially distributed from the physical radius of the planet ($R_{\rm p}$)
to the Roche limit radius (denoted by $a_{\rm R}$), which is given by
\begin{eqnarray}
a_{\rm R} \simeq 2.456\left( \frac{\rho}{\rho_{\rm p}} \right)^{-1/3} R_{\rm p},
\label{roche}
\end{eqnarray}
where $\rho$ and $\rho_{\rm p}$ are the bulk densities of the disk particles and the planet, respectively. 
In this paper, we assume 
$\rho = 0.9\, {\rm {g/cm^3}}$ and $\rho_{\rm p} = 0.7\,{\rm {g/cm^3}}$,
so that $a_{\rm R} \simeq 2.26\,R_{\rm p}$.
The initial surface density of the particles follows $\Sigma(r) \propto r^{-3}$
and their total mass is $\sim (10^{-3}$--$10^{-2}) M_{\rm p}$.
We use $8\times10^4 - 1.2 \times 10^5$ particles with equal mass of $M \sim (10^{-8}$--$10^{-7}) M_{\rm p}$ for the disk.  
Initially, the particles have circular orbits with small enough inclinations, 
following a normal distribution of $\langle e^2 \rangle^{1/2}=2 \langle i^2 \rangle ^{1/2}\sim R/r \sim (2-4) \times 10^{-3}$, where $R$ is the particle physical radius.
Because of the inelastic collisions and self-gravity of the disk particles, they are quickly relaxed to quasi-equilibrium values, which are also $\sim R/r$ ($e$ is a few times larger than $R/r$ probably due to the scattering by the satellite).
The satellite with a mass $M_{\rm s} \sim 10^{-3} M_{\rm p}$ is placed outside the Roche limit.

We do not include the outward orbital migration due to the planetary tide and
eccentricity damping due to the satellite tide,
because they are negligible compared with the migration due to the
satellite-disk interactions at the radius inside the 2:1 resonance with the disk outer edge. 
The tidal $e$-damping and $a$-expansion timescales are 
$\tau_{e, \rm tide} \sim (2/21)(Q_{\rm s}/k_{\rm 2s})(M_{\rm s}/M_{\rm p})(a_{\rm s}/R_{\rm m})^5\Omega^{-1}$ 
and $\tau_{a, \rm tide} \sim (7/2)[(Q_{\rm p}/k_{\rm 2p})/(Q_{\rm s}/k_{\rm 2s})] (M_{\rm p}/M_{\rm s})^{1/3}\tau_{\rm e,tide}$ \citep[e.g.,][]{Charnoz2011}.
For the tidal parameters for the satellite $Q_{\rm s}/k_{\rm 2s}\sim 10^5$, 
$M_{\rm p}/M_{\rm s} \sim 10^6$, and the satellite orbital radius $a_{\rm s}\sim a_{\rm R} \simeq 2.26 R_{\rm p}$, $\tau_{e, \rm tide} \sim 6 \times 10^{10} \Omega^{-1} \sim 2 \times 10^7$ years.
For the planet tidal parameter $Q_{\rm p}/k_{\rm 2p}\sim 10^{3}$--$10^{5}$,
$\tau_{a, \rm tide} \sim 10^7$--$10^9$ years.
As we show in Section 3, the $a$-expansion timescale due to the satellite-disk interactions is
$\tau_{a, \rm disk} \sim (\pi^2/51)(M_{\rm p}/M_{\rm disk})^3(M_{\rm s}/M_{\rm p})(a_{\rm R}/a_{\rm s})^{1/2}\Omega^{-1} \sim  7 \times 10^3 \Omega^{-1} \sim 2 \,{\rm years}$ 
for a realistic case with $M_{\rm disk}/M_{\rm p} \sim 3 \times 10^{-4}$ and $M_{\rm s}/M_{\rm p}\sim10^{-6}$ (see Section 4).
Since $\tau_{a, \rm disk} \ll  \tau_{a, \rm tide}, \tau_{e, \rm disk}$ 
near the disk outer edge, 
the assumption to neglect the tidal force is justified in our simulation. 

Table \ref{tab:param} shows the parameter sets of the runs
with the different initial disk 
mass ($M_{\rm disk}$) and satellite mass ($M_{\rm p}$).
The disk particles have equal masses.
We use the satellite masses that are much larger than the current Saturn's mid-size moons. 
We will derive semi-analytical formulas from the results of N-body simulations to clarify intrinsic physics in this system. 
Applying the derived mass scaling law for the realistic masses of Saturn'{}s mid-size moons, 
we will discuss the possibility to avoid the resonance trapping.

\renewcommand{\arraystretch}{1}
\begin{table}[h]
\begin{center}
\small
\begin{tabular}{c|c|c|c|c|c}
RUN & $M_{\rm disk} [M_{\rm p}]$ & $M_{\rm s} [M_{\rm p}]$ & $N$ & $M_{\rm s,final} [M_{\rm p}]$ & $M_{\rm s,mig} [M_{\rm p}]$ \\ \hline\hline%& $C$\\ \hline\hline
1 & $4.47\times10^{-3}$ & $10^{-3}$  &  $8\times 10^{4}$ & $1.37\times10^{-3}$ &$1.18\times10^{-3}$ \\%& 18 \\
2 & $2.99\times10^{-3}$ & $6\times10^{-4}$ & $10^{5}$ & $7.93\times10^{-4}$ & $7.05\times10^{-4}$ \\%& 6 \\
3 & $8.45 \times10^{-3}$ & $10^{-3}$  &  $8\times 10^{4}$ & $1.87\times10^{-3}$ & $1.60\times10^{-3}$ \\%& 5 \\
4 & $2.11\times10^{-3}$ &  $6\times10^{-4}$  &  $1.2\times 10^{5}$ & $7.27\times10^{-4}$ & $6.71\times10^{-4}$ \\%& 10 \\
5 & $5.99\times10^{-3}$ & $10^{-3}$   &  $10^{5}$ & $1.59\times10^{-3}$ & $1.34\times10^{-3}$ \\%& 15 \\
\end{tabular}
\end{center}
\renewcommand{\arraystretch}{1}
\caption{Parameter sets of our simulations. 
$M_{\rm disk}$ is the initial disk mass.
The unit $M_{\rm p}$ is the host planet mass.
$M_{\rm s}$, $M_{\rm s, final}$ and $M_{\rm s, mig}$ are the masses of the satellite
at $t=0$, at the end of simulations, and at the time when the satellite starts
outward migration, respectively.
$N$ is the initial number of the particles in each run.}
\label{tab:param}
\end{table}

\section{Simulation Results}
  \subsection{Ring Structures}
  
\begin{figure*}[ht]
    \centering
 \includegraphics[width=1.0\hsize, bb=0.000000 0.000000 2048.000000 2048.000000]{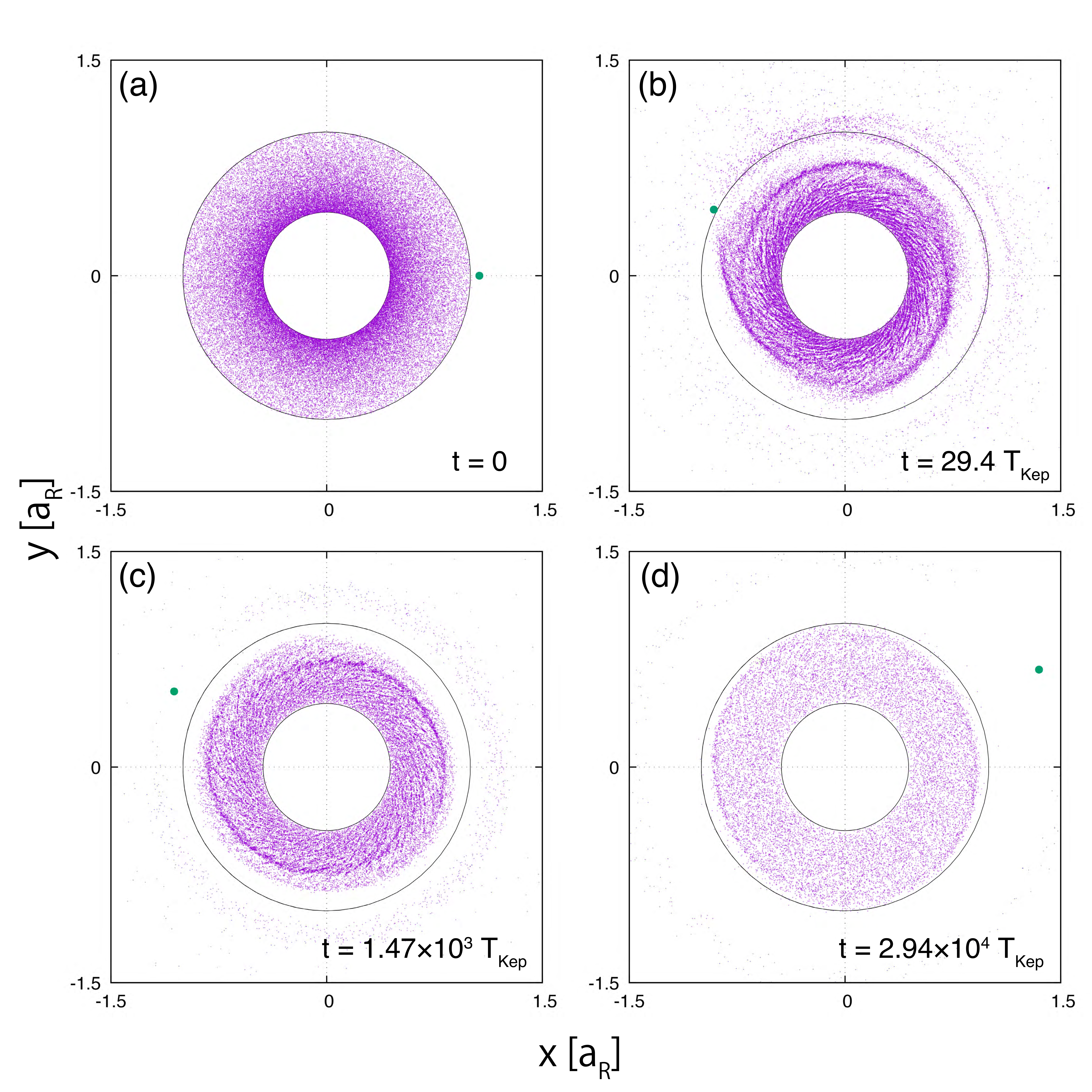}
    \caption{Time evolution of the system of RUN1: (a) $t = 0$, (b) $t = 29.4\,T_{\rm {Kep}}$, (c) $t = 1.47\times10^3\,T_{\rm {Kep}}$ and
  (d) $t = 2.94\times10^4\,T_{\rm {Kep}}$, where $T_{\rm {Kep}}$
  is the Keplerian period at $r=a_{\rm R}$.
  The green point is the outer satellite and purple dots show the disk particles.
   Inner and outer circles in black solid lines represent the planetary surface and
   the Roche limit ($r=a_{\rm R}$).}
  \label{snap_shot}
  \end{figure*}

Figure~\ref{snap_shot}(c) is a snapshot at $t=1.47 \times 10^3 T_{\rm Kep}$ of RUN 1, where $T_{\rm Kep}$ is the Keplerian period at $r=a_{\rm R}$.
The figure shows two two distinct structures are superposed:
1) 
the dense wake structures with small-wavelengths caused by the disk self-gravity 
\citep[e.g.][]{Salo1995,Daisaka2001,Takeda2001}
and 
  2) the $m=2$ spiral arms produced by the Lindblad resonance torque
  from the satellite.
The radial wavelength and wavenumber of the self-gravity wakes are estimated as
\citep{Takeda2001}
    \begin{align}
    \lambda_{\rm self} & \sim 2\pi \frac{M_{\rm disk}}{M_{\rm p}} a_{\rm R}
    \sim 2 \times 10^{-2} \left( \frac{M_{\rm disk}/M_{\rm p}}{3 \times 10^{-3}}\right) a_{\rm R} 
    \label{Lambda} \\
    m_{\rm self} & \sim \frac{2\pi \; a_{\rm R}}{\lambda_{\rm self}}\sim 300
    \left( \frac{M_{\rm disk}/M_{\rm p}}{3 \times 10^{-3}}\right)^{-1}, 
    \end{align}
    where $M_{\rm disk}$ is the total disk mass and we assumed the pitch angle $\sim \pi/4$ to estimate $m_{\rm self}$.
    These estimates are consistent with the result in Fig.~\ref{snap_shot}(c).
Lindblad torque exerted by the satellite makes the spiral arms on the disk \citep[e.g.][]{Goldreich1982}.
The wavenumber of the spiral arms is given by
 \begin{eqnarray}
    m_{\rm res} \sim \frac{\Omega}{\Omega - \Omega_{\rm s}},
\label{arm_res}
\end{eqnarray}
    where $m_{\rm res} \neq 1$ and 
    $\Omega_{\rm s}$ and $\Omega$ are the orbital frequencies of the satellite and the disk.
    Figure~\ref{snap_shot}(c) clearly shows $m_{\rm self}=2$ spiral arms. 
    With $a_{\rm s} \sim 3 R _{\rm p}$ and 
    $\Omega$ at $\sim 2 R_{\rm p}$, Eq.~(\ref{arm_res}) predicts
    $m_{\rm res} \sim 2^{-3/2}/(2^{-3/2} - 3^{-3/2}) \sim 2$, which agrees with Fig.~\ref{snap_shot}(c).
         
Our N-body simulation simultaneously show the short-wavelength wakes due to the self-gravity,
  which were often shown in the local shearing sheet simulations  \citep[e.g.][]{Salo1995,Daisaka2001},
  and the $m_{\rm res}=2$ global spiral arms structure.
While \citet{Hyodo2015} also performed N-body simulation with
 $N=3\times10^4 - 5 \times 10^4$, they did not clearly show 
these two distinct structures, because they used 10 times larger $M_{\rm disk}$ (accordingly,
ten times fewer $m_{\rm res}$) and because their simulations often had multiple clumps.

 \subsection{Time Evolution of Disk structures and the Satellite's orbit}
 
 Figure~\ref{snap_shot}(a) to (d) shows the time evolution of the disk structures.
 We initially set the satellite near the disk outer edge (Fig. \ref{snap_shot}(a)).
 In the disk, the dense wakes quickly emerge due to the
 combined effect of self-gravity and inelastic collisions.
 The satellite creates $m_{\rm res}=3$ spiral arms by the Lindblad torque
 and scatters/accretes nearby disk particles to open a gap with
  the half width $\sim3r_{\rm Hill}$, where $r_{\rm Hill}$ is the satellite Hill's radius defined by $r_{\rm Hill}=(M_{\rm s}/3M_{\rm p})^{1/3} a_{\rm s}$ (Fig. \ref{snap_shot}(b)).
But, the number of the particles scattered outside of the Roche radius is only $\sim 2000$ at this time.  
The disk mass loss is mostly caused by accretion
onto the planet due to the viscous spreading. 
At the initial satellite location ($a_{\rm s} \simeq a_{\rm R}$), 
Eq.~(\ref{arm_res}) at the disk outer edge ($r \sim a_{\rm R}- 3r_{\rm Hill}$) is
\begin{eqnarray}
    m_{\rm res} & \sim &
    \frac{1}{1 - (\Omega_{\rm s}/\Omega)} \sim
    \frac{1}{(3/2) \times 3 (r_{\rm Hill}/a_{\rm s})} \nonumber \\
    &\simeq & \frac{2}{9}\left( \frac{M_{\rm s}}{3M_{\rm p}} \right)^{-1/3} \simeq
     3.2 \left( \frac{M_{\rm s}/M_{\rm p}}{10^{-3}} \right)^{-1/3}.
\end{eqnarray}
This is consistent with the $m_{\rm res}=3$ spiral mode in Fig. \ref{snap_shot} (b).

After the gap opening, the satellite migrates outward, 
and $m_{\rm res}$ decreases from 3 to 2 (Fig. \ref{snap_shot}(c)).   
When the satellite orbit expands beyond the 2:1 resonance with
the disk outer edge, $m_{\rm res}$ becomes well smaller than 2 and
the spiral arms disappears (Fig. \ref{snap_shot}(d)). 
Because the theoretically predicted effective viscosity is 
$\propto \Sigma^2$ (Eq.~(\ref{nu})), 
the $r$-gradient of $\Sigma$ is quickly flattened. 
The self-gravity wakes become fainter (Eq.~(\ref{Lambda}))
through the loss of $M_{\rm disk}$ 
as well as the spiral arms decay (Fig. \ref{snap_shot}(d)).
Consequently, the satellite's orbital migration slows down.
In the next subsection, we quantitatively discuss the satellite migration rate. 
 
 \subsection{Satellite Orbital Evolution}
 
  \begin{figure}[ht]
  \centering
\includegraphics[width=100mm,bb=0 0 200 200]{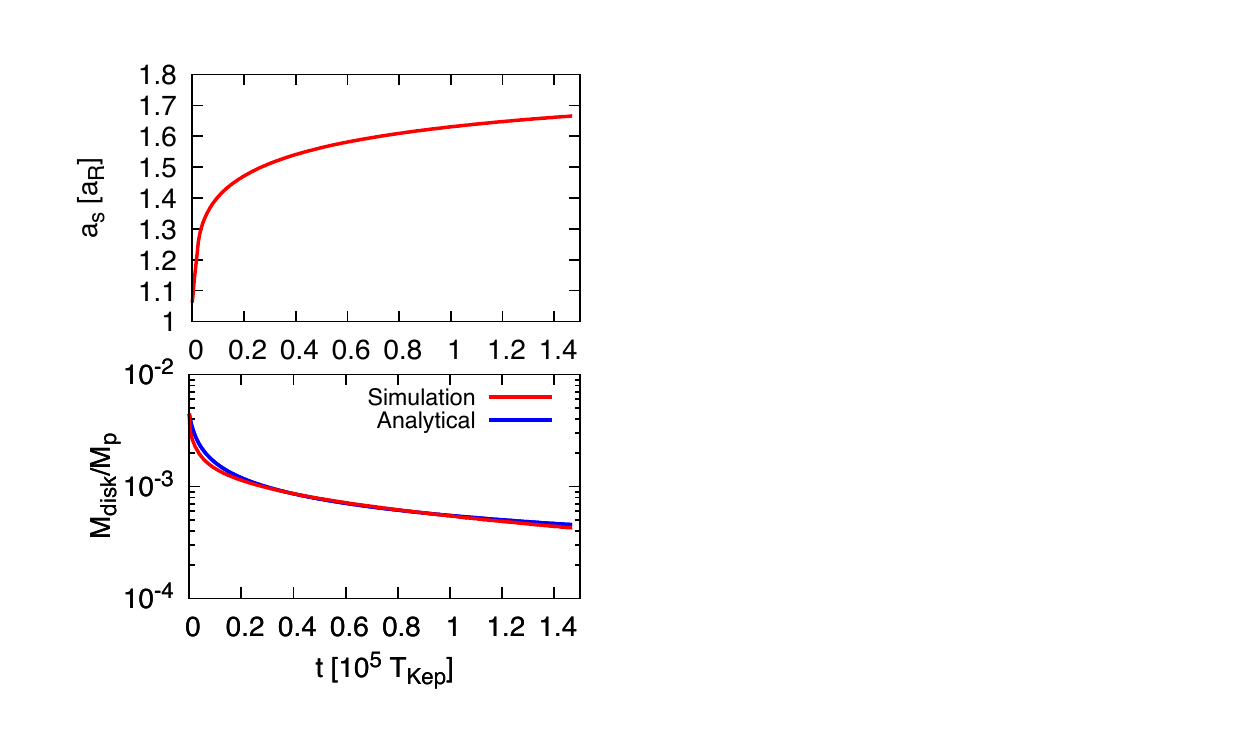}
  \caption{
The time evolution of the satellite's semimajor axis 
($a_{\rm s}$; the upper panel) and the disk mass ($M_{\rm disk}$; the lower panel).
The units of the semimajor axis and the time are the Roche limit radius $a_{\rm R}$ and
the Keplerian period at $a_{\rm R}$, respectively.
The red curves are the results of RUN1 and the blue curve 
in the lower panel is the analytical estimation given by Eq.~(\ref{M_ring}).  
}
  \label{a_evolve}
  \end{figure}
     
Figure~\ref{a_evolve} shows the time evolution
of the satellite's semimajor axis ($a_{\rm s}$) 
and the total disk mass ($M_{\rm disk}$) obtained by our N-body simulation (RUN 1).
The changes of $a_{\rm s}$ and $M_{\rm disk}$ become slower with $t$,
which are theoretically explained as follows.
Through N-body simulations, \citet{Daisaka2001} found that
the effective viscosity of a self-gravitating disk is given by
 \begin{equation}
\nu_{\rm R} \simeq 
26 \gamma
\frac{G^2 \Sigma_{\rm R}^2}{\Omega_{\rm R}}
\simeq \frac{8.5}{\pi^2}\tilde{\gamma}^5 \left( \frac{M_{\rm disk}}{M_{\rm p}}\right)^2 a_{\rm R}^2 \Omega_{\rm R},
  \label{nu}
\end{equation}
where the subscript ''$_{\rm R}$" represents the values at $r \simeq a_{\rm R}$, 
we used $M_{\rm disk} \sim \pi \Sigma_{\rm R} a_{\rm R}^2$, and 
$\tilde{\gamma}=(r/0.8 a_{\rm R})^5$ 
represents the effect of finite physical size of the particles.
Because we are concerned with the outer disk region,
we adopted $r \simeq 0.8 a_{\rm R}$. 
The rate of disk accretion onto the planet is
 \begin{equation}
\dot{M}_{\rm disk} \sim - 3\pi \Sigma_{\rm R} \nu_{\rm R} 
\simeq - \frac{25.5\,\tilde{\gamma}}{\pi^2}
\left( \frac{M_{\rm disk}}{M_{\rm p}}\right)^3 M_{\rm p}\Omega_{\rm R}.
\label{mdot}
\end{equation}
Integrating this equation, we predict the explicit time evolution of the disk mass,
\begin{equation}
\frac{M_{\rm disk}(t)}{M_{\rm p}} \simeq 
\frac{1}{\sqrt{1+(102\,\tilde{\gamma}/\pi) (M_{\rm disk}(0)/M_{\rm p})^2(t/T_{\rm Kep})} }
\frac{M_{\rm disk}(0)}{M_{\rm p}},
\label{M_ring}
\end{equation}
which reproduces the N-body simulation result
(the lower panel of Fig.~\ref{a_evolve}).

Using Eq.~(\ref{M_ring}), we will show that the migration is regulated by
the self-gravity wakes, but not by Lindblad resonance torque (the spiral arms)
induced by the satellite.
The migration rate due to the (one-sided) Lindblad resonance is \citep[e.g.][]{Lin1986,Crida2012}
  \begin{eqnarray}
  \left(\frac{da_{\rm s}}{dt}\right)_{\rm res} & \simeq &
  \frac{16}{27 \pi} \frac{\pi \Sigma_{\rm R} a_{\rm R}^2}{M_{\rm p}}\frac{M_{\rm s}}{M_{\rm p}} 
  \left( \frac{\Delta a_{\rm s}}{a_{\rm R}}\right)^{-3} a_{\rm R}\Omega_{\rm R} \nonumber \\
  & \sim & \frac{16}{27 \pi}\frac{M_{\rm disk}}{M_{\rm p}}\frac{M_{\rm s}}{M_{\rm p}} 
  \left( \frac{\Delta a_{\rm s}}{a_{\rm R}}\right)^{-3} a_{\rm R}\Omega_{\rm R},
  \label{dadt_Lindblad}
  \end{eqnarray}
where $\Delta a_{\rm s} = a_{\rm s} - a_{\rm R}$.
The migration rate by the self-gravity wakes is evaluated as follows.
When the disk viscous spreading beyond the Roche limit is prevented by
the satellite's perturbations, the angular momentum flux in the disk 
($\sim 3 \pi \Sigma \nu \, r^2 \Omega$) is transferred 
from the disk outer edge to the satellite's orbit. 
In this case, 
\begin{eqnarray}
\left(\frac{da_{\rm s}}{dt}\right)_{\rm self} 
 \simeq \frac{2 a_{\rm s}}{L_{\rm s}} \frac{dL_{\rm s}}{dt}
 \simeq \frac{2 a_{\rm s}}{M_{\rm s} a_{\rm s}\Omega_{\rm s}}
 \times 3 \pi \Sigma_{\rm R}  \nu_{\rm R}  \, a_{\rm R}^2 \Omega_{\rm R}.
 \label{dadt_self}
 \end{eqnarray}
Substituting Eq.~(\ref{nu}) and
$M_{\rm disk} \simeq \pi \Sigma_{\rm R} a_{\rm R}^2$
into Eq.~(\ref{dadt_self}), we obatin
\begin{eqnarray}
  \left(\frac{da_{\rm s}}{dt}\right)_{\rm self} 
  \simeq \frac{51 \tilde{\gamma}}{\pi^2}
     \left( \frac{M_{\rm disk}}{M_{\rm p}} \right)^3 
    \frac{M_{\rm p}}{M_{\rm s}} \left(\frac{a_{\rm s}}{a_{\rm R}}\right)^{1/2} a_{\rm R}\Omega_{\rm R}.
   \label{dadt_self2}
  \end{eqnarray}
  
  \begin{figure}[ht]
  \centering
  \includegraphics[width=80mm,bb=30 0 300 200]{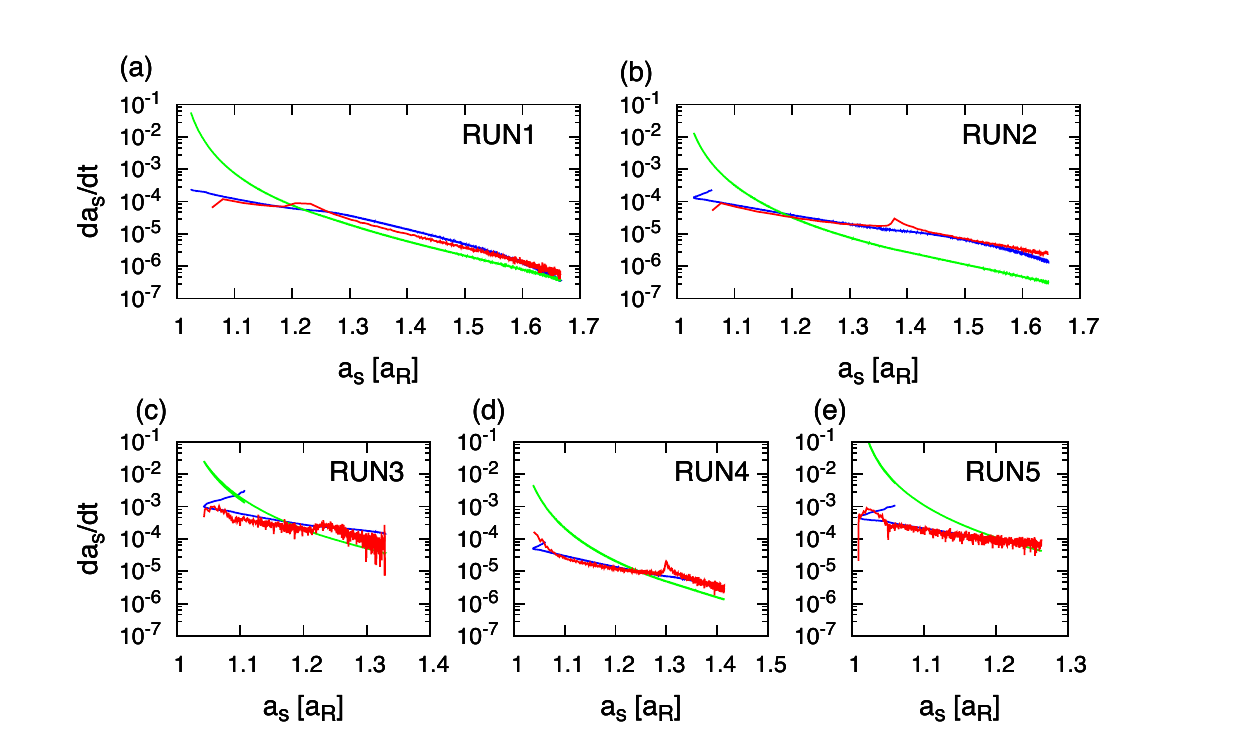}
  \caption{
The orbital expansion rate, $da_{\rm s}/dt$, as a function of $a_{\rm s}$
in RUN1 to RUN5.
The red curves are the results of individual N-body simulations.
The blue and green curves are the theoretical predictions
of $(da_{\rm s}/dt)_{\rm res}$ and 
and $(da_{\rm s}/dt)_{\rm self}$
given respectively by Eqs.~(\ref{dadt_Lindblad})
and (\ref{dadt_self2}) using 
$M_{\rm s}$ and $M_{\rm disk}$ obtained by
the N-body simulations at individual $a_{\rm s}$ in each run.
}
  \label{dadt}
  \end{figure}
 
In Fig.~\ref{dadt}, $da_{\rm s}/dt$ from 
each run of our N-body simulations is compared with
analytical estimations.
In the analytical estimations, 
$M_{\rm s}$ and $M_{\rm disk}$ 
obtained by the N-body simulations at each $a_{\rm s}$
are substituted to Eqs.~(\ref{dadt_Lindblad}) and (\ref{dadt_self2})
to calculate $(da_{\rm s}/dt)_{\rm res}$ 
and $(da_{\rm s}/dt)_{\rm self}$.
This figures show that the results of N-body simulations
fit $(da_{\rm s}/dt)_{\rm self}$.
In the vicinity of the disk edge, $(da_{\rm s}/dt)_{\rm res}$
dominates over $(da_{\rm s}/dt)_{\rm self}$.
The theoretical prediction for $(da_{\rm s}/dt)_{\rm res}$
assumes a non-self-gravitating disk with modest viscosity.
The spiral arms may be weakened by the relatively strong diffusion due to self-gravity wakes.
Because the self-gravity wake torque is independent of the satellite locations,
it dominates over the Lindblad torque that is very sensitive to the distance from the disk outer edge.

In these runs, we adoted $M_{\rm s} \sim 10^{-3}M_{\rm p}$,
while the masses of the actual mid-sized moons are $M_{\rm s} \sim (10^{-7}$--$10^{-6})M_{\rm p}$.
\citet{Hyodo2015} showed through N-body simulations
that $M_{\rm s}/M_{\rm p} \sim 10 (M_{\rm disk}(0)/M_{\rm p})^2$
for $M_{\rm s} \sim 10^{-3}M_{\rm p}$.
Although \citet{Crida2012} proposed
$M_{\rm s}/M_{\rm p} \propto (M_{\rm disk}(0)/M_{\rm p})^3$ for
smaller value of $M_{\rm s}/M_{\rm p}$, generated clumps
would quickly coagulate each other.
Here we use \citet{Hyodo2015}'s relation to discuss the cases of $M_{\rm s} \sim (10^{-7}$--$10^{-6})M_{\rm p}$.
As will be shown in Section 4, the high migration rate 
induced by the self-gravity wake torque, which has been overlooked in the past studies,
would play an important role in avoidance of the mean-motion resonant capture between adjacent satellites.
 
\begin{figure}[ht]
  \centering
\includegraphics[width=100mm,bb=0 0 300 200]{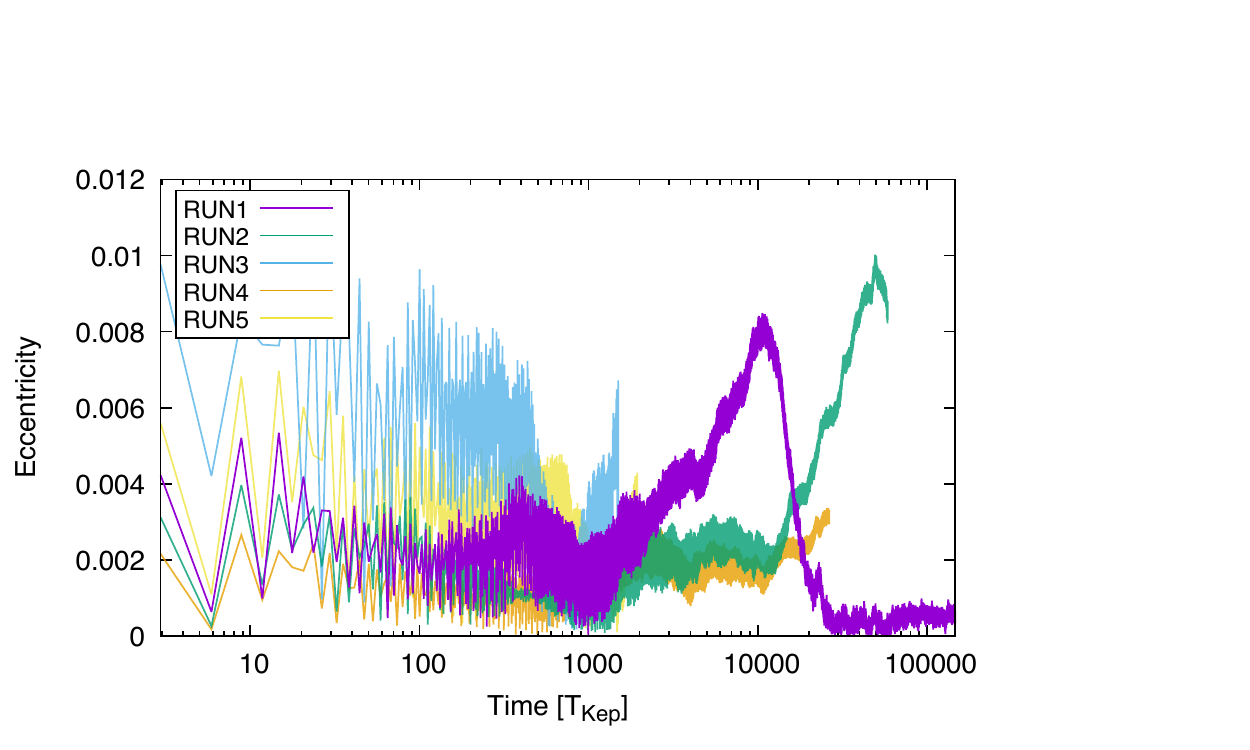}
  \caption{Eccentricity evolution of the massive satellite during the migration in each run of our simulations. 
  }
  \label{e_evolve}
  \end{figure}

  Figure \ref{e_evolve} shows the eccentricity evolution of the satellite
  in the individual runs.
  The eccentricity is excited only in the early phase,
  $t \la (10^{4}$--$10^{5})\, T_{\rm Kep}$,
  when the Lindblad torque may not be negligible compared to 
 the self-gravity wake torque.
  As we discuss in Section 4, the excited eccentricity in the early phase ($e \sim 0.01$)
  is marginal for the condition to avoid a mean-resonance  trapping.   

  \section{Resonance capture probability}
  
   The probability of the mean motion resonance capture 
   depends on the eccentricity and migration rate of the satellites \citep{Dermott1988,Malhotra1993}.
   Here, we consider the avoidance of Tethys-Dione 3:2 resonance tapping as an example.
   Dione's orbit is currently located beyond Tethys' one with the separation within 3:2 mean-motion resonance.
   Because tidal migration rate is a strong function of orbital radius while  
   the mass difference between Dione and Tethys is within a factor of 2  
   ($M_{\rm Dione}/M_{\rm Saturn}\simeq 1.94 \times 10^{-6}, M_{\rm Tethys}/M_{\rm Saturn}\simeq 1.09 \times 10^{-6}$), their tidal migration are convergent unless the planetary tidal parameter $Q_{\rm p}$ 
   is much lower (much more dissipative) for Dione.
   To avoid the trapping into their 3:2 mean-motion resonance,
   a large enough eccentricity and/or fast enough convergence of their orbits is required.

   The critical eccentricity, beyond which the $j+1:j$ resonance trapping is inhibited, is given by
   \citep{Malhotra1996}
    \begin{eqnarray}
    e_{\rm {crit}} 
    \simeq 0.01\left[ \frac{j/(j+1)^2}{0.2}\right]^{1/3}\left( \frac{M_{\rm s}/M_{\rm p}}{10^{-6}}\right).
    \end{eqnarray}  
    \citet{Nakajima2019} pointed out the possibility to avoid Tethys-Dione 3:2 resonance trapping
    by Enceladus' eccentricity excitation.
    If Enceladus' eccentricity is excited enough by the Lindblad torque from the disk,
    Tethys' eccentricity can also be excited to be $\ga e_{\rm crit}$ by the secular perturbations from Enceladus.
        
    The other possibility is fast orbital migration with the timescale 
    shorter than the resonant libration timescale.  
    As we have shown, the migration is significantly faster if we consider the satellite-disk interactions. 
     According to \citet{Ogihara2013}, the resonance trapping is avoided 
     for Tethys-Dione 3:2 resonance ($j=2$), if
    \begin{eqnarray}
    \tau_{a} \equiv \frac{a_{\rm s}}{\dot{a}_{\rm s}} < 
    \tau_{a,{\rm crit}} =  \left(\frac{3}{1024j\alpha^4 f(\alpha)^4} \right)^{1/3} \left( \frac{M_{\rm s}}{M_{\rm p}}\right)^{-4/3} T_{\rm Kep},
    \end{eqnarray}
    where $a_{\rm s}$ is the semimajor axis of the inner satelliite (Tethys),
    $\alpha$ is the semimajor axis ratio ($\simeq 0.763$)
    and $f(\alpha) \sim -1.55/\alpha$ \citep[see][Table 8.5]{MurrayDermott1999}.  
    With $M_{\rm s}/M_{\rm p} \sim 10^{-6}$.
    $\tau_{a,{\rm crit}} \sim 8.0 \times 10^6 T_{\rm Kep}$.

    If we consider the Tethys-mass satellite ($M_{\rm s} \sim 10^{-6} M_{\rm p} $), the disk mass may be 
    $M_{\rm disk} \sim 3 \times 10^{-4} M_{\rm p}$ \citep{Hyodo2015}.
    The orbital migration timescale of the self-gravity wake is estimated to be $\tau_{a,{\rm self}}\sim 1.2 \times10^{3}(M_{\rm disk}/3\times10^{-4}M_{\rm p})^{-3}T_{\rm Kep}$
    (Eq.~(\ref{dadt_self2})),
     which is shorter than $t_{a, \rm crit}$ by more than three orders of magnitude. 
    Although the migration by the self-gravity wake torque could be weaker along with the decrease of $M_{\rm disk}$,
the fast migration potentially prevents the resonance capture of the Dione-Tethys pair. 

\section{Conclusions}

In order to investigate the gravitational interactions between Saturn's mid-sized moons
and a hypothetical ancient massive rings and the associated orbital evolution of the moons, 
we have performed global high-resolution
N-body simulations ($N \sim10^5$) of a self-gravitating particle disk interacting with a single satellite,
taking account of 
gravitational forces among all the disk particles and the satellite
and inelastic collisions between the particles.
Our simulations show that the dense short-wavelength wake structure 
and $m=2$ or 3 global spiral arms simultaneously develop in the disk. 
The former and the latter are produced by the disk self-gravity and 
Lindblad torque from the satellite, respectively.
These structures transfer the angular momentum of the disk to the satellite
and regulate the early phase of orbital evolution of the sattellite.
We found that the orbital migrations of the satellites
are determined by the self-gravity wakes.
The past literatures assumed that the Lindblad torque 
regulates the migrations, because they considered 
the self-gravity wakes only for the source of the disk diffusion.

In this paper, 
we focused on investigating the detailed dynamics 
of gravitational interactions between a circumplanetary particle disk and a satellite, and
derived the semi-analytical formulas for the satellite's migrating rate.
While the simulations used a much more massive satellite than the current Saturn's mid-sized moons due to the simulation limitation, we extrapolated the formulas to realistic satellite masses to find that the migration
is fast enough to avoid the resonance capture of adjacent moons 
on the way to the current orbital configuration of Saturn's mid-size moons.
To confirm this conclusion, the simulations with much higher resolution simulations and with multiple satellites are needed, which is left for future study.

\begin{acknowledgements}
We thank Takaaki Takeda for helpful and useful comments.
This research was supported by JSPS Grants-in-Aid for Scientific Research (\# JP19J12542) and MEXT ``Exploratory Challenge on Post-K computer'' hp190143.

\end{acknowledgements}


\bibliographystyle{aa} 
\bibliography{mybibfile}

\end{document}